# A SIMULATION STUDY FOR T0/T1 DATA REPLICATION AND PRODUCTION ACTIVITIES

**Iosif C. Legrand ***

**Ciprian Mihai Dobre**, Ramiro Voicu**, Corina Stratan**,
Catalin Cirstoiu**, Lucian Musat****

*\* California Institute of Technology*
*\*\* "Politehnica" University of Bucharest*

This paper discusses the latest generation of the MONARC (MOdels of Networked Analysis at Regional Centers) simulation framework, as a design and modeling tool for large scale distributed systems applied to HEP experiments. The simulation of Grid architectures has a vital importance in the future deployment of Grid systems for providing the users an appropriate feed-back. We present here an example of simulating complex data processing systems and the way the framework is used to optimize the overall Grid architecture and/or the policies that govern the Grid's use.

## 1. MONARC SIMULATION FRAMEWORK

The main goal of the MONARC project is to provide a realistic simulation of large distributed computing systems and to offer a flexible and dynamic environment to evaluate the performance of a range of possible data processing architectures. To achieve this purpose, the simulator provides the mechanisms to describe concurrent network traffic and to evaluate different strategies in data replication or in the job scheduling procedures.

A process oriented approach for discrete event simulation is well suited to describe concurrent running programs, network traffic as well as all the stochastic arrival patterns, specific for such type of simulation. Threaded objects or "Active Objects" (having an execution thread, program counter, stack...) allow a natural way to map the specific behavior of distributed data processing into the simulation program.

The simulation program requires the abstraction of all components from the real systems and their time dependent interactions. This abstracted model has to be equivalent to the original system in the key respects that concern us. The simulation engine is designed to be generic and suitable to describe any distributed system. However, there are certain HEP-specific system components that are especially modeled to make the tool useful for the physics community.

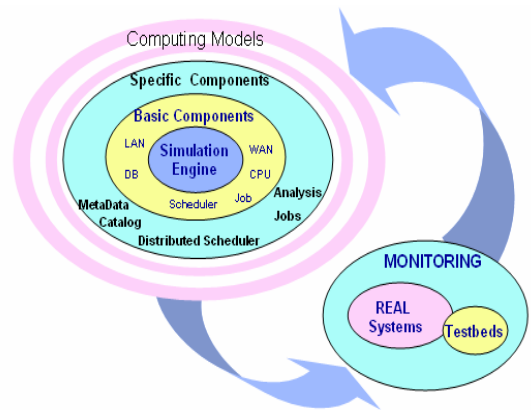

Fig. 1. MONARC 2 layers.

The diagram from figure 1 represents the MONARC 2 layers and the way they could interract with a monitoring system. In the next section we will see one of the simulation studies that we were able to accomplish using the MONARC framework. We simulated the network and computation needs for the distributed activities that will occur in the LHC experiments. The scale, complexity and worldwide geographical spread of the LHC computing and data analysis problems are unprecedented in scientific research. The complexity of processing and accessing this data is increased substantially by the size and global span of the major experiments, combined with the limited wide area network bandwidth available. This simulation study aims to describe the physics analysis processes and the means by which the experiments bands together to meet the technical challenges posed by the storage, access and computing requirements of LHC data analysis.

## 2. A SIMULATION STUDY FOR T0/T1 DATA REPLICATION & PRODUCTION ACTIVITIES

The general concept developed by the two largest experiments, CMS and ATLAS, is a hierarchy of distributed Regional Centers working in close coordination with the main center at CERN. This simulation study follows this concept and describes several major activities; mainly the data transfer on WAN between the T0 (CERN) and a number of several T1 Regional Centers. The topology describing the connectivity of the Regional Centers is presented in figure 2.

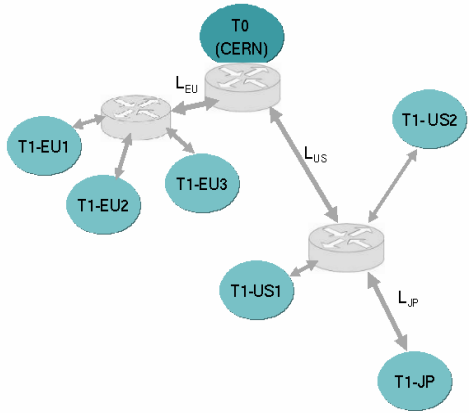

Fig. 2. The network topology considered for the connectivity between the T0 and the T1 Regional Centers.

We assumed that the three T1 Regional Centers in Europe are connected independently, in a network similar to GEANT. In a simplified model this can be approximated with a "mega-router" in which each T1 regional center is connected through a link. We also consider a transatlantic link connecting T0 with the two T1 regional centers in US and another link connecting the T1 regional centers in Japan. In order to make the file transfer efficient we assume that a transfer Agent runs on all the centers. When it is necessary to send a file to several or all of these centers we have assumed that this is done using the Agent mechanism to provide effective data transfers. In case the same file needs to be transferred to both T1 regional centers in US, the file is transferred only once over the transatlantic line and than copied from T1-US1 to T1-US2 or / and T1-JP.

For the WAN links we assumed the following RTT values:

Table 1 RTT values used in the simulation.

| Link | RTT(ms) |
|---|---|
| T1-EU1 <-> T0 (CERN) | 20 |
| T1-EU2 <-> T0 (CERN) | 25 |
| T1-EU3 <-> T0 (CERN) | 30 |
| T1-US1 <-> T0 (CERN) | 120 |
| T1-US1 <-> T1-US2 | 60 |
| T1-US1 <-> T1-JP | 240 |

Those RTT values are used in evaluating the efficiency of using the available bandwidth for "ftp" like transfers. Using this topology we simulated a number of Activities specific for Physics data production, as follows:

1. **RAW Data Replication**. From the experiment we assumed a mean rate of recording raw data equal to 200 MB/s. This information is stored in 2GB (normal distributed with 10% sd) data files. These files are replicated in a round robin manner to all 6 T1 regional centers. (The first file is sent to T1-EU1, the second to T1-EU2…)

2. **Production and DST distribution**. At T0 all raw data are processed and DST files are generated. The DST files are 10 times smaller in size than the RAW files. We considered again a normal distribution (sd 10%). The DST files created at T0 are sent to all T1 centers. For the T1-US2 and T1-JP the agent transfer system is used to make this operation effective and avoid sending the same file more than once over the same link.

3. **Re-production and new DST distribution**. After a certain time the RAW data in each T1 center is re-processed and new DST data is created. Each T1 center will reprocesses 1/6 of the RAW data. The DST data generated at each regional center are sent to all the other. Again the agent system is used to effectively transfer data.

4. **Detector Analysis**. This activity starts in certain T1 regional centers at given moments of time and collects all RAW data from the other regional centers produced over the last hours. We chose local 9 o'clock as the time this activity starts in the given regional centers and also we chose to gather the RAW data for the last 12 hours. The RAW data is gathered dynamically, meaning from all the regional centers that have the requested data it is chosen the one that maximizes the performance of the transfer, based on the network load, proximity and database load.

We simulated the described activities alone and then combined. We simulated approximately one day of running these activities. In the following figures are some conclusions obtained when running all four activities in parallel. We assumed a mean rate of recording raw data of 200 MB/s. The information is stored in 2GB data files (normally distributed with 10% sd). DST files are produced in the second activities involved at T0 (CERN) from all the RAW data and then are distributed to all the T1 regional centers. The data transfer agent described above is then used. After a certain period of time each T1 center will start to re-process the raw data stored locally and to generate a new set of DST. Each T1 has ~1/6 from the entire raw data and will generate new DST which should be replicated to all the other regional centers. As before, in this case we also assume that transfer agents are running on all the centers involved (T0, T1-US1) for an effective replication. Finally, the Detector Analysis activity runs on T1-JP regional center and starts at 9 o'clock local time. Then it will gather the RAW data produced in the last 12 hours from the others centers using a get-optimum-performance algorithm as mentioned above.

Using this configuration we did a series of tests in which we have varied the available bandwidth between T0 (CERN) and T1-US1. In the following figures are the obtained results.

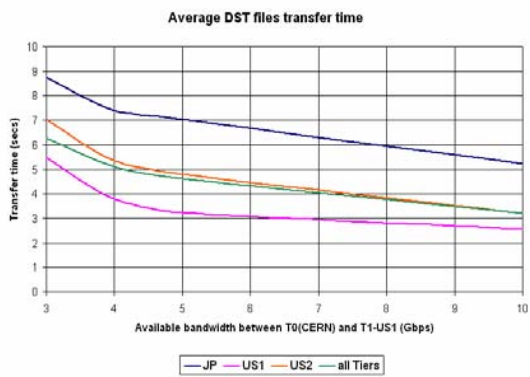

Fig. 3. The DST files transfer time in different T1 centers with different values for the available bandwidth between T0 (CERN) and T1-US1.

In the figure 3 is the representation of how varies the time with which the DST files are served in different T1 centers for the test cases in which the available bandwidth between T0 (CERN) and T1-US1 varies between 3Gbps and 10Gbps. As seen the DST files transfer time tends to decrease proportionally with the amount of bandwidth available between T0 (CERN) and T1-US1 centers. The series "all Series" represents the average value of the DST files transfer time considering all the T1 tiers in the simulation.

In figure 4 is the representation of the way the RAW file transfer time varies in different T1 centers in the tests in which we have varied the amount of available bandwidth between T0 (CERN) and T1-US1. As seen the RAW files transfer time tends to decrease proportionally with the amount of bandwidth available between T0 (CERN) and T1-US1 centers. The series "all Series" represents the average value of the RAW files transfer time considering all the T1 tiers in the simulation.

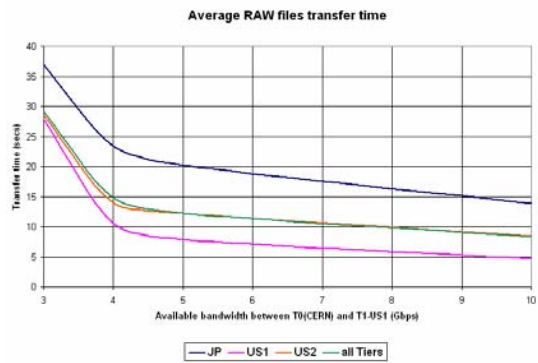

Fig. 4. The RAW files transfer time in different T1 centers with different values for the available bandwidth between T0 (CERN) and T1-US1.

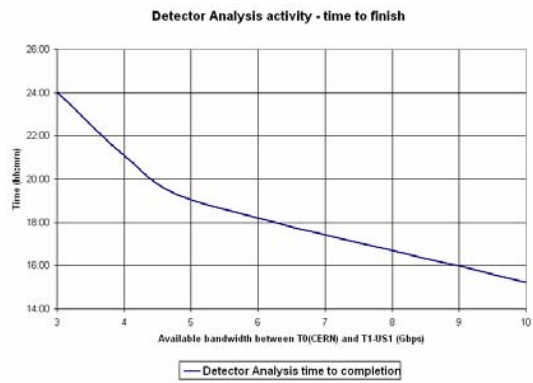

Fig. 5. Time needed for the Detector Analysis activity to finish for the tests done centers with different values for the available bandwidth between T0 (CERN) and T1-US1.

In the figure 5 is the representation of the variation of time needed to complete the Detector Analysis activity in the tests done for different values for the amount of available bandwidth between T0 (CERN) and T1-US1. As said above this activity gathers the RAW data from the last 12 hours, but as seen here when using a 3Gbps link it takes almost 24 hours to finish, while when using a 10Gbps link between T0 (CERN) and T1-US1 it takes around 15 hours to finish.

We will present as follows a comparison for production and DST distribution, done with and without the Data Transfer Agent.

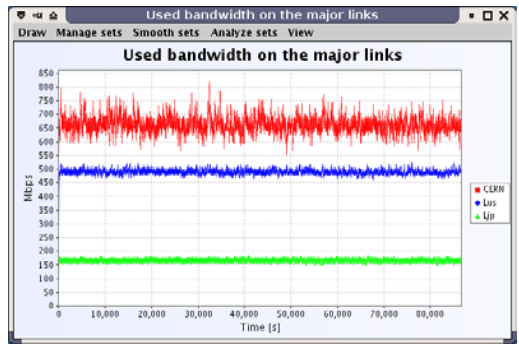

Fig. 6. The used bandwidth on the major links output obtained for the test done using the Data Transfer Agent.

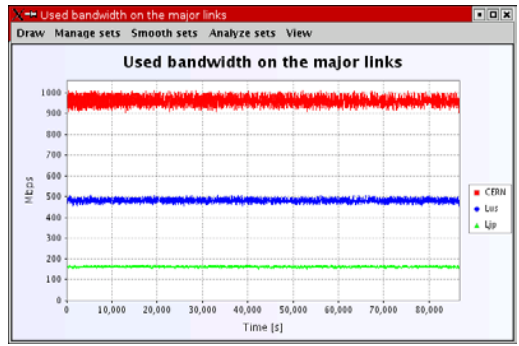

Fig. 7. The used bandwidth on the major links output obtained for the test done without using the Data Transfer Agent.

In the Production and DST distribution activity test at T0 (CERN) regional center are produced DST files from the recorded RAW data, which are then distributed to all the T1 regional centers. The Data Transfer Agent is used on the T1-US1 regional center and will forward the DST data received in that center from T0 (CERN) further to T1-US2 and T1-JP regional centers (see figure 1). This means that at T0 (CERN) when using the Data Transfer Data the DST files will be sent only to T1-EU regional centers and to T1-US1, while the agent will handle the further transfer of those files from T1-US1 to the rest of the regional centers. For the average used bandwidth on the major links the obtained results are shown in figures 6 and 7. As seen the average bandwidth used on the CERN link is greater when we do not use the Data Transfer Agent since more data get transferred from that regional center.

Furthermore, the difference between the two tests is better seen in the figures 8 and 9, which show the results obtained for the parameter average used bandwidth again on the major links, but on each direction on each link (from CERN to EU, from CERN to US1, etc).

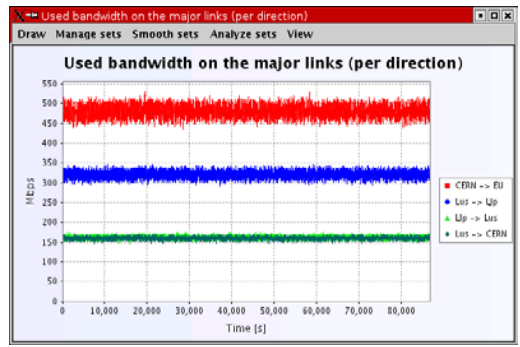

Fig. 8. The used bandwidth on the major links on each direction output obtained for the test done using the Data Transfer Agent.

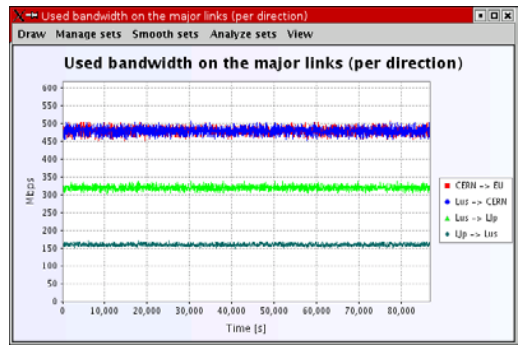

Fig. 9. The used bandwidth on the major links on each direction output obtained for the test done without using the Data Transfer Agent.

The results are explained if we look into the deeper details of the tests done. In both tests at T0 (CERN) regional center are produced DST files with a continuous rate. The desired purpose in this test is, as already mentioned, to distribute those DST files from this regional center to all the other regional centers. One of the tests uses a Data Transfer Agent on T1-US1 that will transfer further the files to T1-US2 and T1-JP. So, in both tests from T0 (CERN) to T1-EU is transferred the same amount of data (see link CERN->EU in the graphics below). But the difference consists in the amount of data that is transferred from T0 (CERN) to T1-US1.

When using the Data Transfer Agent only one file gets transferred on that connection for all three regional centers (the T1-US1 will act as a proxy for that file), while in the test done without using the

Data Transfer Agent one file will be transferred through that link to each of the three regional centers. In the graphics below this is represented by the link Lus->CERN and as seen the report is 1:3 as expected in the bandwidth that gets used. The link Lus->Ljp represents the connection from T1-US1 to the other two regional centers, and in both tests two files are transferred through it. Ljp->Lus represents the link that arrives at T1-JP, so again in both tests only one file gets transferred each time through it.

In conclusion we have found out that using a Transfer Agent in the hub T1 centers is important to save resources for the data replication activities. Also for the assumed values, a 2.5Gbps link from T0 (CERN) to US is not enough to keep up with the traffic generated by the production activity.